# Properties of the Broad-Range Nematic Phase of a Laterally Linked H-Shaped Liquid Crystal Dimer

Young-Ki Kim,[a] Randall Breckon,[b] Saonti Chakraborty,[c] Min Gao,[a] Samuel N. Sprunt,[c] James T. Gleeson,[c] Robert J. Twieg,[b] Antal Jákli,[a] and Oleg D. Lavrentovich[a]*

[a] *Liquid Crystal Institute and Chemical Physics Interdisciplinary Program, Kent State University, Kent, OH 44242, USA*
[b] *Department of Chemistry & Biochemistry, Kent State University, Kent, OH 44242, USA*
[c] *Department of Physics, Kent State University, Kent, OH 44242, USA.*

* Corresponding author. E-Mail : olavrent@kent.edu

**Abstract**

In search for novel nematic materials, a laterally linked H-shaped liquid crystal dimer have been synthesized and characterized. The distinct feature of the material is a very broad temperature range (about 50 $^o$C) of the nematic phase, which is in contrast with other reported H-dimers that show predominantly smectic phases. The material exhibits interesting textural features at the scale of nanometers (presence of smectic clusters) and at the macroscopic scales. Namely, at a certain temperature, the flat samples of the material show occurrence of domain walls. These domain walls are caused by the surface anchoring transition and separate regions with differently tilted director. Both above and below this transition temperature the material represents a uniaxial nematic, as confirmed by the studies of defects in flat samples and samples with colloidal inclusions, freely suspended drops, X-ray diffraction and transmission electron microscopy.

# 1. Introduction

Liquid crystals (LCs) with their numerous thermodynamically stable phases present a dramatic illustration of how a small alteration of the molecular structure can lead to profound changes in the long-range order. The simplest uniaxial nematic phase that enabled the LCD industry, is typically formed by rod-like molecules with a straight central rigid core and two aliphatic chains. The rods prefer to align parallel to each other, along a single axis, called the director $\hat{n}$, setting anisotropic character of all physical properties of the material.[1] An intriguing example is represented by molecules with two rod-like rigid elements connected by a flexible aliphatic tail with an odd number of methylene groups [2-10]: these dimers form the so-called twist-bent nematic phase [11-14] with nanoscale periodic modulation of the director.[15, 16] As shown by transmission electron microscopy (TEM), the underlying nanoscale periodic motif represents an oblique helicoid.[16] The structure is thus intermediate between the normal uniaxial nematic $(N_u)$ and the chiral nematic.

The fourth nematic is the so-called biaxial nematic $(N_b)$, in which the uniform alignment of, say, the longest axes of the molecules, is accompanied by uniform orientation of the short axes. $N_b$ was first identified by Yu and Saupe in a lyotropic mixture of potassium laurate / 1-decanol / water system.[17] Although the biaxial order has been observed by other researchers in this and other lyotropic nematics,[18-27] the issue remains controversial, as the studies by Berejnov et al. [28, 29] suggest that the biaxial order is only transient.

Existence of a thermotropic $N_b$ is even less clear. The most recent search focused on the so-called bent-core mesogens. In the bent-core materials with oxadiazole units, $N_b$ was identified in the studies of X-ray diffraction (XRD),[30, 31]

NMR & conoscopy,[32] electro-optical switching,[33-36] and topological defects.[37] However, reexamination of how materials behave in the external fields and in confined geometries with topological defects, lead to a conclusion that two oxadiazole compounds, abbreviated C7 and C12, are uniaxial nematics that mimic the behavior of $N_b$ because of effects such as surface anchoring transitions.[38, 39]

The $N_b$ phase was also claimed in the azo-substituted bent-core nematic A131, following XRD and DSC,[40] NMR,[41] conoscopy,[42, 43] and Raman scattering experiments.[44]. More recent NMR studies resulted in a conclusion that it is necessary to test the material for possible conformational changes [45] and that a transition within the nematic range of A131, identified as one from the uniaxial to the biaxial phase, can be associated with a slowing down of the molecular rotations around the long molecular axis.[46] Two other similar azo compounds have also been claimed to feature the $N_b$ phase on the basis of XRD [40, 47] and conoscopy [43] studies. However, explorations of electro- and magneto-optics, surface alignment and defects showed that the nematic order in A131 is uniaxial.[39, 48, 49]

One of the reasons for the controversy is that uniaxial nematic phase can mimic the features of the $N_b$ phase in a variety of forms. Very often this mimicking behavior is rooted in the complexity of surface alignment of molecules with nontrivial shapes such as bent-core [38, 49] and tetrapodes mesogens.[50] An interesting example of mimicking behavior is demonstrated by a mesogen with four lateral flexible chains.[51] The biaxial-like features occur as a result of thermal expansion which triggers flows; the latter cause a biaxial tilt of the uniaxial director that can persist for hours because of high viscosity of the material.[51, 52]

The examples above are by no means an indication that all of the potentially $N_b$

are in fact uniaxial. As of today, the claimed existence of the $N_b$ phase has not been challenged for bent-core mesogens with oxadiazole [31, 53, 54] and azo-groups,[40, 43, 47] for strongly asymmetric units,[55, 56] and for the shape-persistent derivatives of benzodithiophene [57, 58] and fluorenone.[43, 59-61] The state of the art of biaxiality in bent-core nematics and in other shape LCs have been reviewer recently.[62, 63] Other molecular geometries, such as mesogens with four flexible chains,[64, 65] tetrapode derivatives [66-71] different from the one explored in Ref. [50] also show $N_b$ features that have not been disputed so far. Finally, there is an interesting but underexplored class of compounds with cyclic (ring-like) mesogens synthesized by Percec [72-74] with the $N_b$ features in phase behavior [72, 73] and in structure of wall defects.[74]

It is of interest to note that optical biaxiality in the uniaxial nematic can be induced by the external electric or magnetic field, including the case of bent-core mesogens [75-80]. However, the field-induced effects involve multiple mechanisms, including uniaxial and biaxial modifications of the order parameter and also modification of the director fluctuations, which are often difficult to separate from each other.[81, 82] In this work, we deal with the situations in which the electric field suppresses potentially biaxial features (such as domain walls, see below for more details) and thus does not induce the $N_b$ phase.

Obviously, the different symmetry of the long-range nematic order implies different types of physical properties and ensuing applications. Uniaxial nematics are widely exploited in LCDs, chiral nematics are used in light-reflecting devices, while twist-bend nematics show interesting polar electro-optic effects [4, 10] and distinct electric field-induced reorientation caused by dielectric anisotropy.[16] The biaxial nematics are expected to speed up the electrooptic response as compared to their uniaxial counterparts.[83] The short review presented above makes it clear, however, that the identification of the true connection between the molecular structure and phase symmetry is often a difficult task and requires one to use multiple experimental techniques, especially those that do not depend on the integrated response of the sample but instead probe the local order, as in the case of topological defects.

In this work, in search for novel nematic materials, we have synthesized (Synthetic processes are available in Electronic Supplementary Information) and explored the LC abbreviated RB01189, formed by mesogenic dimers in the shape of letter 'H', the so-called H-dimers, Figure 1a, b. Typically, H-dimers form smectic phases.[84-88] The distinct feature of the material explored in this work is a very broad temperature range of the nematic phase, about 50 degrees. The material shows intriguing textural features, both at the scale of microns and at the scale on nanometers. In particular, optical textures reveal peculiar defects that appear at a certain temperature. A detailed study of the optical features of these defects through regular polarizing microscopy and quantitative mapping of optical retardance demonstrates that they represent domain walls that appear as a result of surface anchoring transition. The walls separate regions with different tilt of the director $\hat{n}$. Both above and below the temperature at which the domain walls appear, the H-dimer material retains its uniaxial nematic order. At the scale of nanometers, the material show traces of smectic clustering.

## 2. Results and discussion

### 2.1 Phase diagram, alignment, and birefringence

To establish the phase diagram and to explore the textures of RB01189, we used flat glass cells with transparent indium tin oxide electrodes. The cell thickness $d$ was set by spacers. All the experiments were performed above the crystallization temperature to prevent a possible memory effect.[38] The temperature was controlled by a hot stage LTS350 with a controller TMS94 (both Linkam Instruments) with $0.01^{o}$C accuracy. A typical rate of temperature change was $\pm 0.1^{o}$C/min to minimize the effects of thermal expansion.[51, 52] The phase diagram of RB01189 determined by polarizing microscopy upon cooling is:

$$\mathrm{Cr} < 56^{o}\,\mathrm{C} < \mathrm{N} < 108^{o}\,\mathrm{C} < \mathrm{I}$$

where Cr, N, and I indicate the crystal, nematic, and isotropic phases, respectively. The N-I transition temperature $T_{NI}$ varies within $1-3^{o}$C depending on the cell thickness $d$ and alignment layer; $T_{NI}$ is constant within the entire area of the same cell. The material shows a $2^{o}$C biphasic region with coexisting N and I phases.

We used two polyimides as the aligning layers (AL), namely, PI2555 (AL1, HD Microsystems) and SE5661 (AL2, Nissan Chemical Industries). For the classical uniaxial LCs such as 5CB and E7, AL1 provides a tangential (in-plane) alignment, while AL2 yields a perpendicular (homeotropic) alignment of the director $\hat{\mathbf{n}}$. In the case of RB01189, the situation is more complex. Judged by the Schlieren textures with isolated centers showing two and four extinction brushes (disclinations and point defects, respectively), AL1 provides a strictly tangential anchoring in the entire temperature range of the N phase, there is no tilt.[1, 89] AL2 shows either a tangential

Accepted in Liquid Crystals

or tilted alignment of $\hat{\mathbf{n}}$, depending on temperature, as discussed in the next section. We used unidirectionally rubbed cells of thickness $d = 1.4\,\mu\text{m}$ with AL1 to measure the birefringence $\Delta n$ as a function of the effective temperature $t = T - T_{\text{NI}}$ in the N phase, Figure 1c.

**2.2 Anchoring transitions and domain walls.**

AL2 cells show an interesting sequence of textures that depend both on temperature and the cell thickness $d$. At high temperatures, the cell with $d = 6\,\mu\text{m}$ shows Schlieren textures typical of tangentially anchored cells, Figure 2a, with defect centers at which two or four extinction bands converge. Below a certain temperature $t_1$, a network of domain walls (DWs) with interference colors different from the background emerges in the texture, Figure 2b. Their appearance is very similar to the so-called "secondary disclinations" identified as the disclinations that signal presence of $N_b$.[37] However, in our case, these defects represent DWs of the uniaxial nematic, associated with the tilted surface anchoring at the substrates of the $N_u$ cell, as analyzed below.

As the temperature decreases, the contrast of the DWs first enhances and then decreases, Figure 2c-f. To quantify the change in the contrast, we used PolScope technique that maps the optical retardance $\Gamma$ of the cell as a function of the in-plane $(x, y)$ coordinates.[90-92] The maximum value of the measured retardance allowed by PolScope is limited by 273 nm, thus we used a thin cell with $d = 1.4\,\mu\text{m}$, Figure 3. In the thin cell, the DWs form immediately after the isotropic-nematic transition, i.e., $t_1 = 0°\text{C}$. The optical retardance at the center of the DW is higher than in the background, Figure 3a-d. As the temperature decreases, the background retardance increases and eventually, the retardance of the surrounding regions matches the retardance of the DW. Variation of $\Gamma$ along a line that crosses one of the DWs (dotted line in Figure 3a-d) is presented in Figure 3e for different temperatures. The maximum value of $\Gamma$ ($\Gamma_{max}$) is reached at the center of the DW (labeled by a symbol * in Figure 3a-d). For an extended temperature range, $\Gamma_{max}$ matches the maximum possible value of retardance $d\Delta n$ that is reached when the director is strictly planar, Figure 3f. The fact that $\Gamma_{max} = d\Delta n$ means that the defect represents a DW [89] with a tangentially aligned central part, separating regions with tilted director and thus smaller retardance. In Schlieren textures of samples with tilted alignment, the DWs are stabilized by the defect centers with two extinction brushes, since these centers correspond to vertical disclinations around which the director rotates by $180°$. In contrast, the cores with four extinction brushes that correspond to the point defects-boojums [89] can accommodate the titled alignment without creating DWs. In Figures 2d,e, one can clearly see that the centers with two brushes are connected by DWs but the centers with four brushes remain DW-free.

Figure 4 shows the detailed director pattern and the retardance map around DWs in an AL2 cell of $d = 1.1 \mu m$ at $t = -1.4 °C$. The center of the DW is easily identifiable as the region with the highest retardance. One can distinguish two limiting cases, depending on the predominant deformation of the director $\hat{\mathbf{n}}$ across the wall: a splay-bend DW, Figure 4b, and a twist DW, Figure 4c. Note also that the DWs might also occur as closed loops, some of which are visible in Figure 2b,c.

To prove further that the textural defects in AL2 cells are DWs caused by the tilted surface alignment, we explored the electrooptic response of the texture. RB01189 has a negative dielectric anisotropy, $\Delta\varepsilon < 0$. Applying an electric field across the cell, one should reorient the director perpendicular to the field and thus establish a tangential orientation everywhere with no DWs. Figure 5 shows that this is indeed the case. By applying a vertical AC electric field (sinusoidal wave, frequency $f = 1\,\text{Khz}$) to the AL2 cell with $d = 1.4\,\mu\text{m}$, one increases the retardance in the regions outside the DWs, which implies that the director becomes more parallel to the bounding plates. The maximum retardance measured at $5\,\text{V}_{\text{PP}}$ is 93 nm. For $d = 1.4\,\mu\text{m}$, it corresponds to the effective birefringence 0.066, which is close to the birefringence measured 0.064 in the planar cell at the same temperature, $t = -1.5^\circ\,\text{C}$. At sufficiently high voltage $(> 4\,\text{V}_{\text{PP}})$, the DWs disappear and the retardance does not show any significant variation in the plane of the sample (except at the cores of disclinations and boojums at which the retardance is smaller).

We conclude that the observed DWs are caused by the tilted alignment in AL2 cells. In thick samples, the orientation is tangential above $t_1$ and tilted below $t_1$. As the director realigns from the tangential to tilted, the DWs form in order to separate regions with the different direction of tilt. The fact that these DWs are not observed in AL1 cells (in which the director remains parallel to the bounding plates in the entire range of the N phase) and the fact that the temperatures of appearance of DWs in thin and thick AL2 cells are different from each other, further supports the conclusion that the DW are not associated with the $N_u - N_b$ transition. The next section confirms the $N_u$ type of ordering in the entire range of the RB01189 nematic phase, through the studies of defects in colloidal-LC composites.

## 2.3. Point defects in droplets and around colloidal spheres

Topological defects are often used for phase identification of LCs, as these are uniquely defined by the order parameter type.[89] The straightforward approach [38, 49, 50, 93-95] is to use colloidal settings in which the topological defects correspond to the equilibrium state of the LC. It can be a spherical droplet of a LC that must feature a certain number of topological defects because of its shape,[93] or it can be a colloidal sphere immersed in the LC bulk that causes similar defects around itself.[96] We explored both situations for RB01189, Figure 6.

RB01189 droplets were dispersed in glycerol (Sigma-Aldrich). They show a bipolar texture with two isolated boojums in the entire N range, Figure 6a, d-g. Such a texture is a classic example of a uniaxial nematic confined in the sphere with tangential surface anchoring.[93, 97] It is very different from the biaxial nematic droplet, in which the boojums at the poles must be connected by a disclination line as shown in Figure 6b, or combine two boojums into one, Figure 6c.

We have also explored the surface point defects-boojums around spherical colloids that set a tangential orientation of the surrounding director in RB01189. The borosilicate glass spheres were dispersed in uniformly aligned planar AL1 cells of $d = 20\,\mu m$. Each spheres produces only a pair of boojums at its poles and no other defects, Figure 6 j,k, which is again consistent with the uniaxial (Figure 6h) but not with the biaxial charater (Figure 6i) of order, as discussed in Refs. [38, 49]. The features of topological defects in colloid-LC samples confirm that RB01189 is a uniaxial nematic.

## 2.4. Smectic nanoclusters in the nematic phase : small angle X-ray scattering study & TEM observation.

The small angle X-ray scattering (SAXS) patterns with four peaks are similar to those for the uniaxial nematic phase of bent-core LCs containing smectic C clusters, i. e., show nanoinclusions with layered structure in which the molecules are tilted with respect to the normal to the layers. [98, 99] The distance of the maxima of the four spots from the origin yields the wavenumber $q = 0.285\,\text{Å}^{-1}$ corresponding to a layer periodicity of $p = 2.2\,\text{nm}$, which is considerably shorter than the approximate length $l \sim 3.7\,\text{nm}$ of the fully extended aromatic legs of the dimers, indicating a large tilt angle $\theta$ that the aromatic legs make with the smectic layers' normal. The temperature dependence of the tilt angle $\theta$, determined from the angle between the magnetic field and the maxima of the spots, is shown in Figure 7a. The tilt angle is indeed large, over $50^\circ$, and barely depends on temperature below the biphasic range (more than $2^\circ\text{C}$ away from the isotropic phase). The full width at half maxima of the $I(q)$ curves yields $\Delta q \sim 0.13\,\text{Å}^{-1}$, which is also fairly constant in the entire nematic range. The correlation length $\xi$ of the smectic clusters is therefore $\xi = 2/\Delta q \sim 1.54\,\text{nm}$, which corresponds to a typical cluster size of about $3.1\,\text{nm}$ (somewhat increases toward lower temperatures). The short correlation length and the small cluster size imply that the smectic layers in the LC bulk are poorly correlated. The reason might be the possibility of interdigitated arrangements of the H dimers on both sides of a single smectic layer with a large tilt angle $\theta$. These interdigitated arrangements might involve multiple conformational states with different shifts of the molecules along the direction of aromatic legs and thus would weaken the smectic correlations.

Direct Cryo-TEM (figure 7b) and Freeze-Fracture TEM (Figure 7c) observations were carried out following the sample preparation described in Ref. [16, 100] to check the physical presence of layered clusters. Most areas do not show periodic structures, which is consistent with the small size of the cybotatic groups. However, in some areas, we find regions with periodic modulations that extend in the direction perpendicular to the "layers" over a significant distance of about 100 nm. The periodicity of the fringes is (3.4-3.8 nm) in both cryo-TEM, Fig.7b, and in freeze-fracture TEM, Fig.7c, which is close to the length of the aromatic leg of the dimer, but significantly larger than the smectic layer spacing deduced from SAXS experiments. The cryo-TEM data might imply that the very small thickness of the cryo-TEM specimen (typically less than 100 nm) forces the molecules to be in the plane of the sample and reduces the tilt angle $\theta$. A similar effect of confinement-induced suppression of the molecular tilt has been observed in smectic phases formed by bent-core molecules.[101] The freeze-fracture TEM data might be explained by the tendency of fractures to be along the lowest density of molecules, which is the direction parallel to the longest extension of the molecules.

## 3. Conclusions

In this study, we synthesized an H-shaped LC dimer RB01189 with a broad-range nematic phase and explored its physical properties. The material is a uniaxial nematic with embedded smectic nanoclusters that exist in the entire range of the nematic phase. The smectic nanoclusters are evident not only in the X-ray scattering experiments, but are also directly visible in two different modes of TEM observations.

Despite the complex molecular structure, the rubbed polyimide layers of PI2555 can align RB01189 uniformly. At some aligning substrates, such as SE5661, the H-dimer material shows surface anchoring transitions, with the director changing its surface orientation from tangential to tilted. The transition is accompanied by appearance of the domain walls, across which the uniaxial director changes its orientation from tilted to planar and back to tilted, to accommodate two neighbouring domains with a different direction of tilt. The walls are similar in appearance to the secondary disclinations described recently as a feature of the biaxial order but in the H-material under study, they have a very different nature associated with the variation of surface orientation of the uniaxial director. Finally, the broad temperature range of the nematic phase makes the H-dimers suitable candidates for the search of the remaining missing nematic, the so-called splay-bend phase.[11, 12, 14] It is expected that H-dimers with an odd number of methylene groups in the aliphatic bridge might form splayed configurations that would facilitate a formation of the splay-bend nematic. These studies are in progress.

**4. Experimental section**

**Sample Preparation for Small Angle X-ray Scattering Studies:** The materials were filled into 1 mm diameter quartz X-ray tubes, which were then mounted into a custom-built aluminium cassette that allowed X-ray detection with $\pm 13.5^\circ$ angular range. The cassette fits into a standard hot stage (Instec model HCS402) that allowed temperature control with $\pm 0.01^\circ$ C precision. The stage also included two cylindrical neodymium iron boron magnets that supplied a magnetic induction of $H = 1.5\,\text{T}$ at the position of the X-ray capillary. The magnetic field direction is perpendicular to the incident X-ray beam. Two-dimensional SAXS images were recorded on a Princeton Instruments $2084 \times 2084$ pixel array CCD detector in the X6B beamline at the National Synchrotron Light Source. The beamline was configured for a collimated beam $(0.2\,\text{mm} \times 0.3\,\text{mm})$ at energy $16\,\text{keV}\,(0.775\,\text{Å})$. Details of the experimental conditions are described elsewhere.[98]

## Acknowledgements

This work was supported by the DOE under Grant DE-FG02-06ER 46331 (optical, TEM, and polarizing microscopy studies) and the NSF under Grant DMR-0964765 and DMR-1307674 (XRD studies).

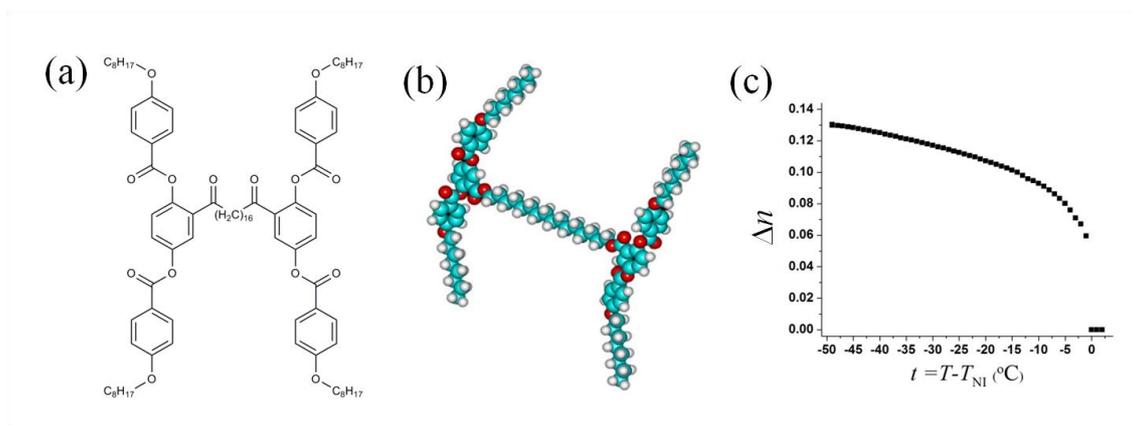

Figure 1. (a) Chemical structure and (b) space filling model of H-shaped RB01189 molecules. (c) Birefringence of RB01189 measured in the planar AL1 cell as a function of effective $t = T - T_{\text{NI}}$.

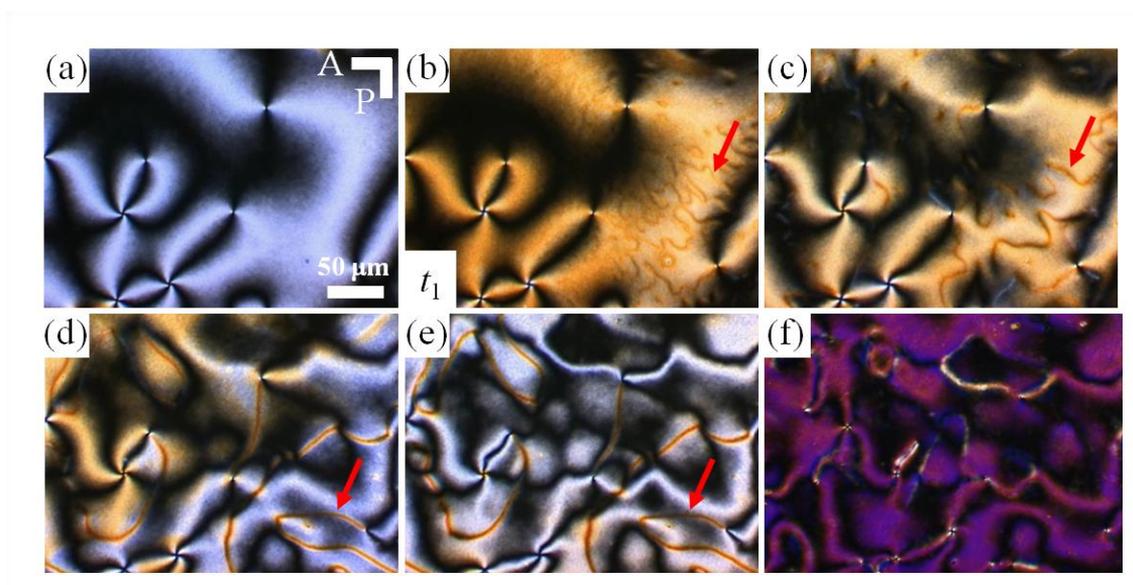

Figure 2. (Colour on-line) Sequence of textures in the AL2 cell ($d = 6$ μm) at (a) $t = -1°\text{C}\,(T = 107°\text{C})$, (b) $-1.8°\text{C}\,(= t_1)$, (c) $-2°\text{C}$, (d) $-2.3°\text{C}$, (e) $-2.5°\text{C}$, and (f) $-4°\text{C}$; A and P indicate analyzer and polarizer, respectively; arrows point towards the domain walls.

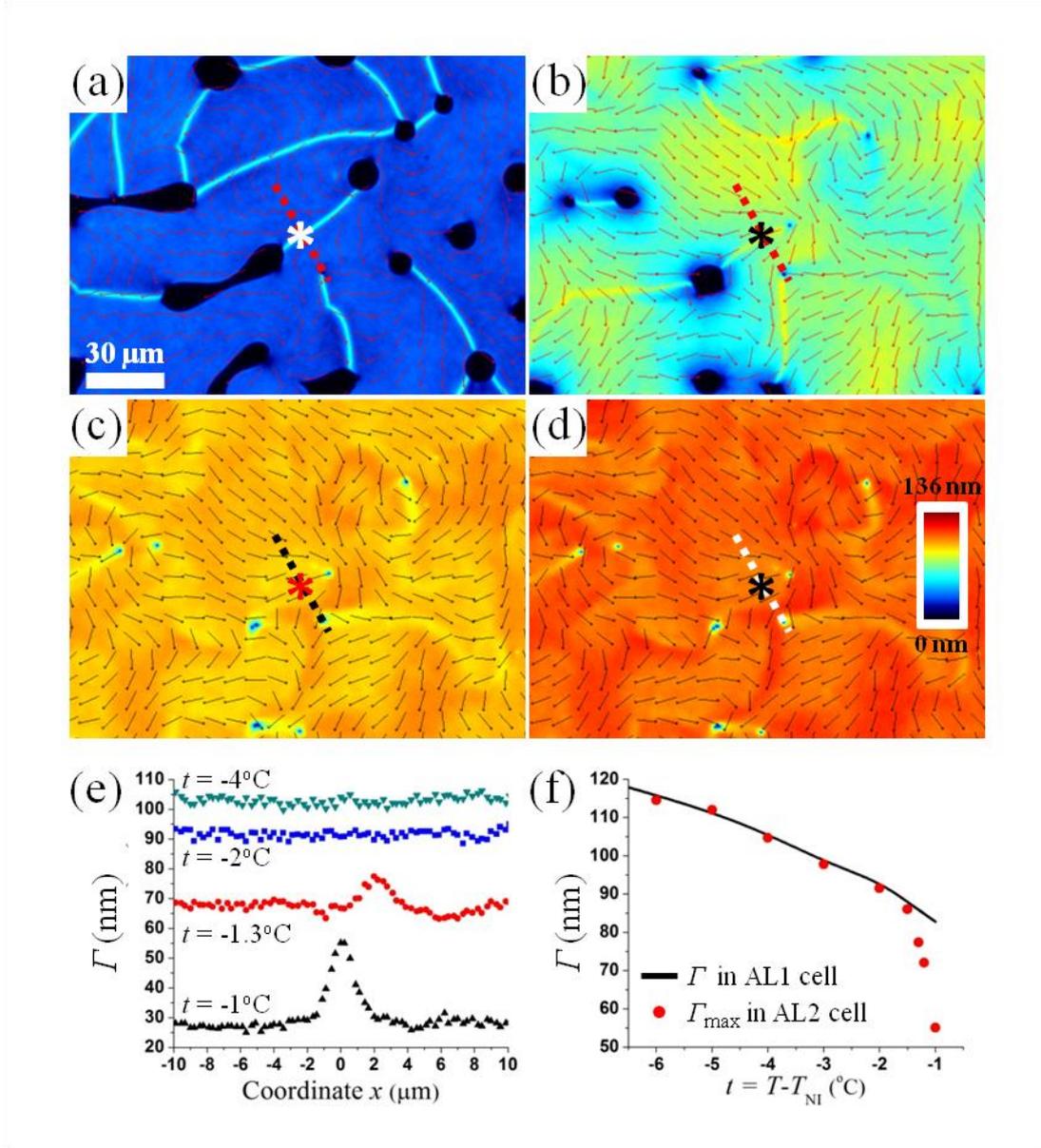

Figure 3. (Colour on-line) Retardation maps in the AL2 cell ($d = 1.4$ μm) at (a) $t = -1°\text{C}(<t_1)$, (b) $-1.3°\text{C}$, (c) $-2°\text{C}$, and (d) $-4°\text{C}$; grids represent the director $\hat{\mathbf{n}}$. (e) Measured $\Gamma$ along the dotted lines in a-d. (f) Comparison of $\Gamma = d\Delta n$ for $d = 1.4$ μm ($\Delta n$ data are measured experimentally in planar cell, Figure 1c) and $\Gamma_{max}$ (measured at the center of the DW marked * in a-d) in the AL2 cell of thickness $d = 1.4$ μm.

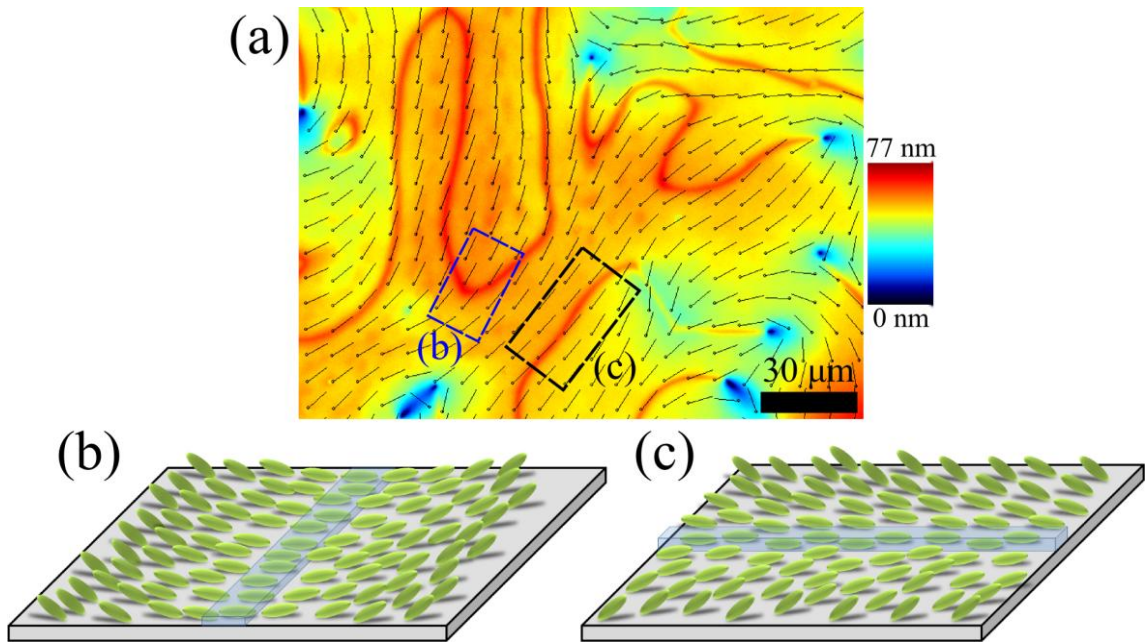

Figure 4. (Colour on-line) (a) Retardation map with the director field projected onto the plane of the cell in the AL2 cell ($d = 1.1$ μm) at $t = -1.4°\text{C}(< t_1)$; black grids indicate the director $\hat{\mathbf{n}}$. Director configurations for (b) bend-splay and (c) twist DW.

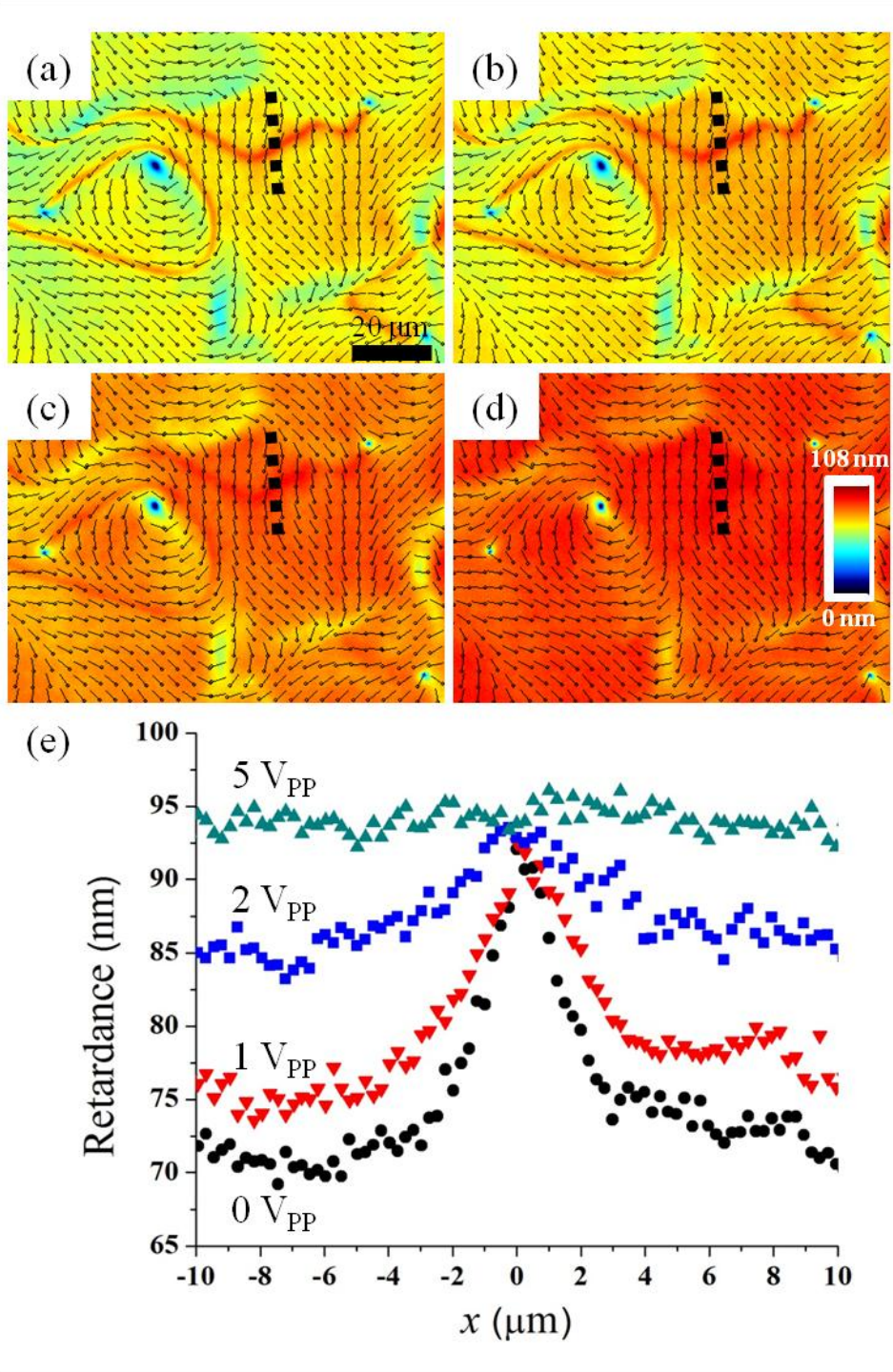

Figure 5. (Colour on-line) Retardation maps of the AL2 cell ($d = 1.4$ μm) at $t = -1.5°$ C with the vertical electric field of (a) 0, (b) 1, (c) 2, and (d) $5\,V_{PP}$; grids map the director $\hat{\mathbf{n}}$. (e) Retardance across DW (along dashed lines).

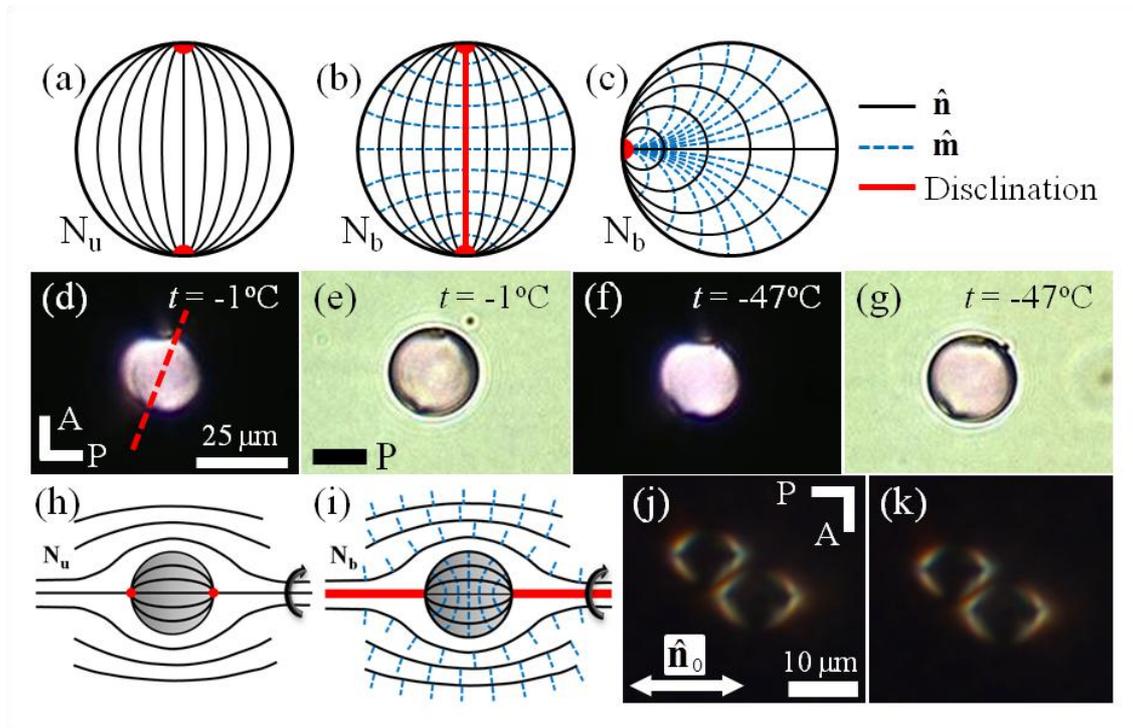

Figure 6. (Colour on-line) (a) $N_u$ bipolar droplet with two point defect-boojums. $N_b$ bipolar droplets (b) with a singular disclination $m=1$ (red bold line) formed by secondary director $\hat{\mathbf{m}}$ and (c) with a single boojum of strength $m=2$. (d-g) $N_u$ bipolar droplet of RB01189 in a glycerol at (d, e) $t=-1^\circ\text{C}\,(T=107^\circ\text{C})$, and (f, g) $-47^\circ\text{C}$; dashed line indicates the droplet symmetry axis. (h-k) Scheme of director configuration around colloids in (h) $N_u$ and (i) $N_b$. Chain of colloids with the $N_u$ bipolar structure in AL1 cell ($d=20$ μm) at (j) $t=-1^\circ\text{C}$ and (k) $-47^\circ\text{C}$.

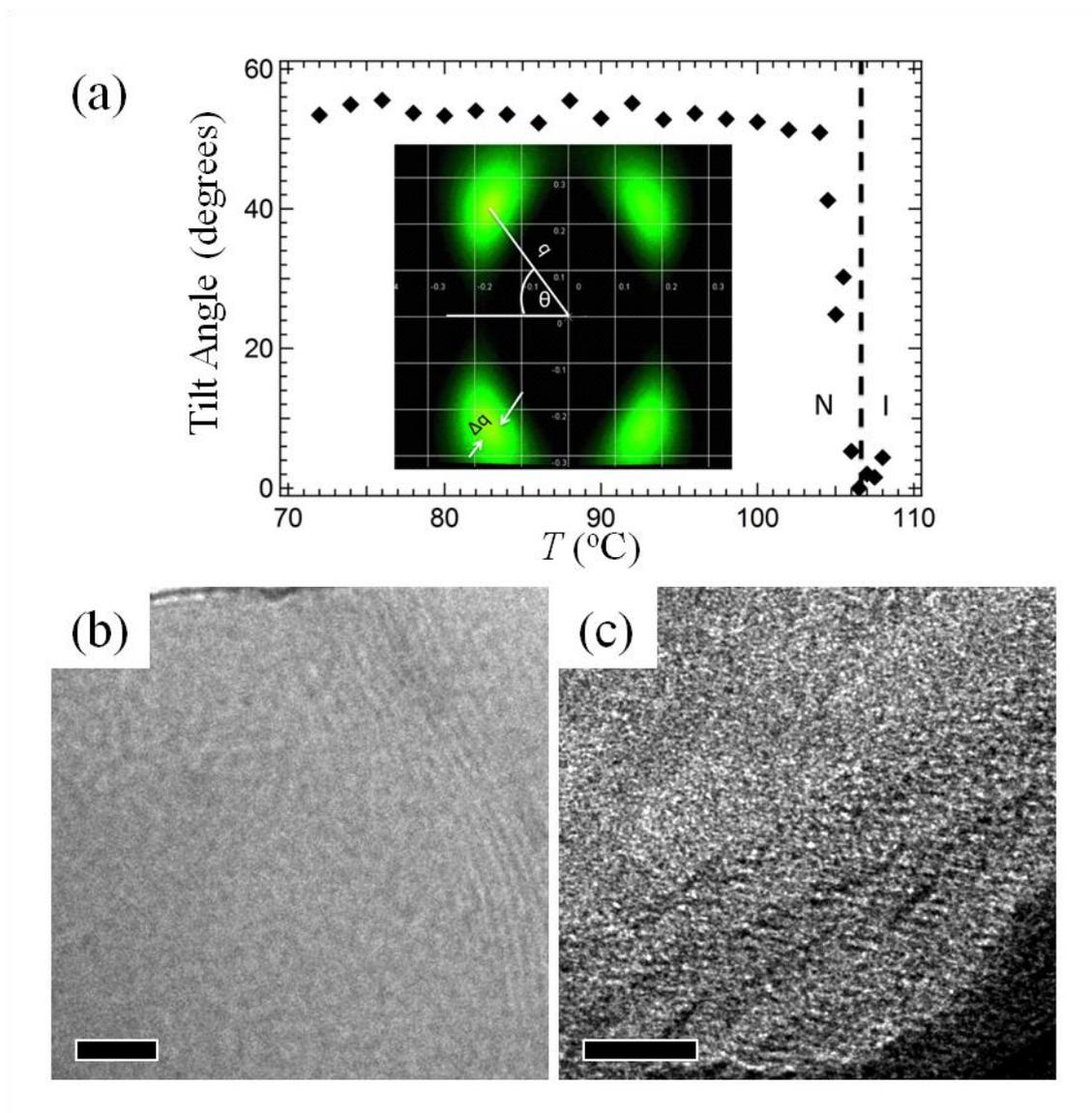

Figure 7. (a) Temperature dependence of the tilt angle of the smectic clusters in the N phase of RB01189. The inset shows SAXS pattern at $T = 75^\circ$ C. (b,c) Direct images of the smectic nanoclusters taken by (b) Cryo-TEM and (c) freeze-fracture TEM; sample was quenched at $75^\circ$ C. Scale bar is 25 nm.

# Properties of the Nematic Phase of a Laterally Linked H-Shaped Liquid Crystal Dimer

*Young-Ki Kim,[a] Randall Breckon,[b] Saonti Chakraborty,[b] Min Gao,[a] Samuel N. Sprunt,[c] James T. Gleeson,[c] Robert J. Twieg,[b] Antal Jákli,[a] Oleg D. Lavrentovich[a]*

[a] *Liquid Crystal Institute and Chemical Physics Interdisciplinary Program, Kent State University, Kent, OH 44242, USA.*
[b] *Department of Physics, Kent State University, Kent, OH 44242, USA.*
[c] *Department of Chemistry & Biochemistry, Kent State University, Kent, OH 44242, USA.*

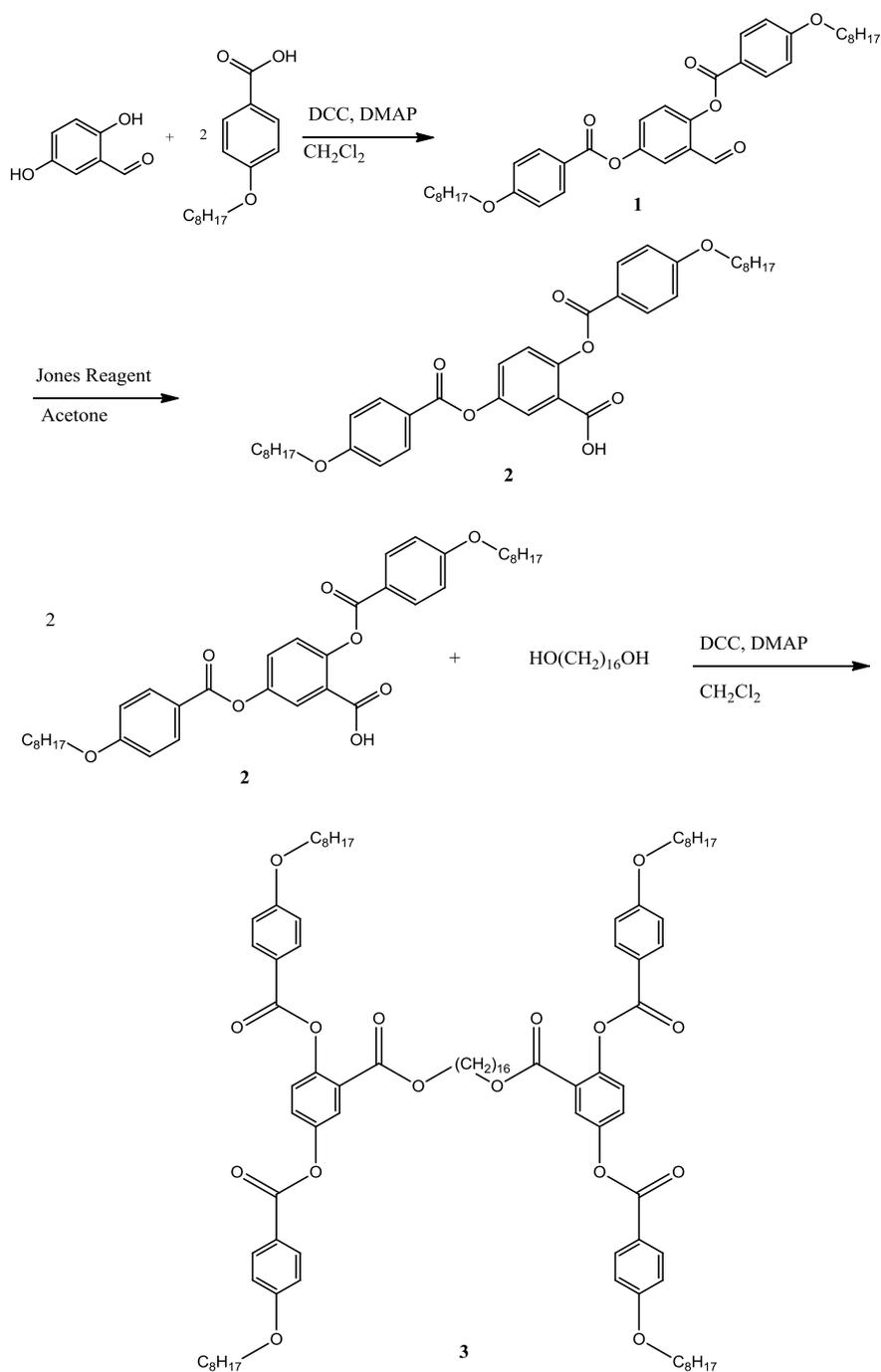

**Experimental**

DSC's were obtained on a TA Instruments Modulated 2920 DSC at rate of 5°C/min under an $N_2$ atmosphere. The $^1$H NMR were obtained on a Bruker Avance 400MHz NMR using TMS as an internal standard and $^{13}$C NMR were also obtained on this instrument at 100MHz using TMS as the internal standard.

*2-formyl-4-{[4-(octyloxy)phenyl]carbonyloxy}phenyl 4-(octyloxy)benzoate* (**1**). To a solution of 3.00g (21.7 mmol) of 2,5-dihydroxybenzaldehyde and 13.57g (54.3 mmol) of 4-octyloxybenzoic acid dissolved in 200ml of $CH_2Cl_2$ was added 11.19g (54.3 mmol) of DCC and 0.080g (0.64 mmol) of DMAP. This solution was stirred at room temperature overnight. The resulting mixture was vacuum filtered through Celite to remove the DCU byproduct. The solvent was removed *in vacuo* and the resulting solid was recrystallized from ethanol to afford the product as an off-white powder. Yield 10.585g (81.0%). $^1$H NMR (CDCl$_3$), δ 10.24 s (1H), 8.20 m (4H), 7.81 d (1H, *J*= 2.8 Hz), 7.56 dd (1H, *J*= 2.8, 8.8 Hz), 7.41 d (1H, *J*= 8.8 Hz), 7.02 m (4H), 4.09 m (4H), 1.86 m (4H), 1.51 m (4H), 1.33 m (16H), 0.92 t (6H, *J*= 6.8 Hz). $^{13}$C NMR (CDCl$_3$), δ: 187.5, 164.6, 164.5, 164.1, 163.8, 150.0, 148.8, 132.6, 132.4, 129.1, 128.8, 124.7, 122.3, 120.8, 120.3, 114.6, 114.4, 68.4, 31.8, 29.3, 29.2, 29.1, 26.0, 22.7, 14.1. M.P. 150-152 °C.

*2,5-bis({[4-(octyloxy)phenyl]carbonyloxy})benzoic acid* (**2**). To a solution of aldehyde **1** dissolved in 200ml of acetone was added 17.1 ml (26.4 mmol) of 1.54M Jones Reagent.[1] This solution was stirred overnight at room temperature. The solvent was then removed *in vacuo* and 500ml of water were added to the remaining solids. The resulting slurry was stirred for one hour and the solids were filtered off and washed copiously with water to give the product as a white powder. Yield 10.812g (99.4%). $^1$H NMR (CDCl$_3$), δ 8.15 m (4H), 7.96 d (1H, *J*= 2.4 Hz), 7.53 dd (1H, *J*= 2.8, 8.8 Hz), 7.32 d (1H, *J*= 8.8 Hz), 6.99 m (4H), 4.08 m (4H), 1.85 m (4H), 1.51 m (4H), 1.33 m (16H), 0.92 t (6H, *J*= 6.4 Hz). $^{13}$C NMR (CDCl$_3$), δ: 164.9, 164.5, 163.8, 163.6, 148.7, 148.3, 132.5, 132.4, 128.0, 125.5, 125.2, 123.4, 121.3, 120.9, 114.4, 114.3, 68.4, 68.3, 31.8, 29.3, 29.2, 29.1, 26.0, 22.7, 14.1. M.P 167-169 °C.

**Exemplary method for the synthesis of H-shaped adducts from carboxylic acid 2:**

*16-{[2,5-bis({[4-(octyloxy)phenyl]carbonyloxy})phenyl]carbonyloxy}hexadecyl 2,5-bis({[4-(octyloxy) phenyl]carbonyloxy})benzoate* (**3**). To a solution of 0.500g (0.81 mmol) of acid **2** and 0.101g (0.39 mmol) of 1.16-hexadecanediol that were dissolved in 25 ml $CH_2Cl_2$ was added 0.167g (0.81 mmol) of DCC and 0.010g (0.08 mmol) of DMAP. This solution was stirred at room temperature overnight. The resulting mixture was filtered through Celite to remove the DCU byproduct. The solvent was removed *in vacuo* and the resulting solid was recrystallized from 3-methyl-1-butanol to afford a white powder. Yield 0.329g (65.5 %). IR $\nu_{max}$ (Neat)/cm$^{-1}$ 2921, 2852, 1731, 1606, 1510, 1248, 1160, 1063, 760. $^1$H NMR (CDCl$_3$), δ 8.18 m (8H), 7.92 d (2H, *J*= 2.8 Hz), 7.48 dd (2H, *J*= 2.8, 8.8 Hz), 7.28 d (2H, *J*= 8.4 Hz), 7.00 m (8H), 4.17 t (4H, *J*= 7.2 Hz), 4.07 m (8H), 1.85 m (8H), 1.50 m (12H), 1.33 m (32H), 1.25 m (24H), 0.92 m (12H). $^{13}$C NMR (CDCl$_3$), δ: 164.9, 164.6, 164.2, 163.8, 163.6, 148.3, 148.1, 132.5, 132.4, 127.0, 125.0, 121.5, 121.0, 114.4, 114.3, 68.4, 68.3, 65.7, 31.8, 29.7, 29.6, 29.5, 29.3, 29.2, 29.1, 28.4, 26.0, 25.9, 22.7, 14.1.

DSC    Heating Cycle- **Cr** 101.3 **N** 105.3 Iso
         Cooling Cycle- Iso 103.9 **N** 31.6 **Cr**